**Astro2010 Science White Paper (GCT)**
*Fundamental Accretion and Ejection Astrophysics*
J. Miller, M. Nowak, P. Nandra, N. Brandt, G. Matt, M. Cappi, G. Risaliti, S. Kitamoto,
F. Paerels. M. Watson, R. Smith, M. Weisskopf, Y. Terashima, Y. Ueda

**Accretion Disks, Winds, and Jets**

Disk accretion may be *the* fundamental astrophysical process. Stars and planets form through the accretion of gas in a disk. Black holes and galaxies co-evolve through efficient disk accretion onto the central supermassive black hole. Indeed, approximately 20% of the ionizing radiation in the universe is supplied by disk accretion onto black holes. And large-scale structures – galaxy clusters – are dramatically affected by the relativistic jets that result from accretion onto black holes. Yet, we are still searching for *observational* answers to some very basic questions that underlie all aspects of the feedback between black holes and their host galaxies:

- **How do disks transfer angular momentum to deliver gas onto compact objects?**
- **How do accretion disks launch winds and jets?**

X-rays probe disk accretion directly, in the region closest to the black hole itself. Leveraging this advantage into strong physical tests of accretion mechanisms, however, requires the ability to obtain excellent spectra on relevant orbital and even sub-orbital timescales. This, in turn, requires a combination of high spectral resolution and a large collecting area – the X-ray equivalent of the Keck telescopes or ESO Very Large Telescope. The jump in collecting area for high resolution spectroscopy from *Chandra* to IXO exceeds the jump from 2-4 meter to 8-10 meter optical telescopes. *With IXO, astrophysicists will be able to study the physics that drives disk accretion, and connections between disks, winds, and jets.*

Optical spectroscopy reveals that disks around pre-main sequence stars transfer matter onto the star, and angular momentum outward, partially though a magneto-centrifugal wind (Blandford & Payne 1982; Calvet, Hartmann, & Kenyon 1993). This means that poloidal magnetic field lines extend from the accretion disk, and that gas escapes along the field lines as the disk rotates. Far from the disk, such magneto-centrifugal winds may collimate into a jet (Blandford & Payne 1982). Moreover, the basic field structure necessary for such winds is essential for tapping the spin energy of the black hole itself to power Poynting jets (Blandford & Znajek 1977). There is strong *theoretical* support for common physics driving accretion onto pre-main sequence stars, and onto compact objects. At present, however, there are only tantalizing observational hints of this commonality (Mauche & Raymond 2000; Kraemer et al. 2005; Miller et al. 2006).

*Chandra* observations of Seyfert AGN and some quasars reveal disk winds in X-ray absorption, sometimes called "warm absorbers" (see, e.g., Kaspi et al. 2002, Blustin et al. 2005, Krongold et al. 2007). Observations of NGC 1365 have already sampled one distinct transit event (Risaliti et al. 2005, 2009). With IXO, absorption due to orbiting material will be seen to change as an absorber transits the face of the central engine. For instance, an absorber in a Keplerian orbit at 1000 $GM/c^2$ will take more than 30 ksec to cross the face of a central engine with a radius of 20 $GM/c^2$, for a $10^7$ solar mass black hole. During the transit, absorption lines will shift by 400 km/s in total. Based on the X-ray absorption currently observed, IXO will detect a typical H-like O VIII line (for instance) and determine its velocity shift to high accuracy in just 5 ksec (see Figure 1), thus revealing the orbital transit. *The ability to observe orbital motion in disk-driven outflows will be revolutionary, placing black hole accretion on the same footing as stellar accretion and revealing details of angular momentum transfer, wind and jet production.*

*To understand disk accretion is to understand a core aspect of feedback and the co-evolution of black holes and their host galaxies* (e.g. Begelman, de Kool, & Sikora 1991, Silk & Rees 1998, Ciotti & Ostriker 2007). The disk winds seen in Seyfert AGN and some quasars are expected to move up to $10^8$ solar masses of material during their active lifetimes (Blustin et al. 2005), and the main component of the mass and kinetic luminosity is only visible in the X-ray band. This gas amounts to the dominant source of hot interstellar material in many galactic bulges. Moreover, surveys show that AGN tend to show evidence for recent episodes of star formation (Kauffmann et al. 2003), and outflows from AGN may contribute to this process through shocks (Crenshaw & Kraemer 2000).

Accretion onto stellar-mass black holes and neutron stars provides an important point of comparison. Although single transits cannot be observed with IXO, for sources as bright as $10^{-9}$ erg/cm$^2$/s, detailed absorption spectra can be obtained on timescales corresponding to Keplerian orbits at just 200 GM/c$^2$. Changes in disk winds can be tracked precisely in stellar-mass systems, including black holes and neutron stars. Lines are stronger in some white dwarf systems, meaning that lower flux levels can be reached.

The fundamental accretion exploration described here can be conducted in at least 20 AGN. Generous integrations of 100 ksec each will only require 2 Msec – a low fraction of the IXO observing budget over 5 years. (This work can be done *simultaneously* with explorations of General Relativity using relativistic disk lines.) In a 5-year mission, 5 transient stellar-mass black holes, 5 transient neutron stars, 30 persistent neutron stars, and 30 white dwarfs can be studied. With modest integrations of 30 ksec, only 2.1 Msec will be required to build a stellar-mass sample for comparison with AGN and optical/IR observations of pre-main sequence stars.

It is worth noting that accretion studies with IXO will tap a deep discovery space, fortuitously supported by a golden age in accretion theory. Simulations of accretion disks and outflows are advancing rapidly, as computational power grows and becomes widely available. Around the world, a rapidly increasing number of research groups are producing disk simulations that will test analytical work and make new predictions (see, e.g., Balbus & Hawley 1991, Miller & Stone 2000, Proga 2003, Gammie 2004, Anderson et al. 2005, Blaes et al. 2006, Hawley, Beckwith, & Krolik 2007, McKinney & Blandford 2008). The crystal spectrometer that flew aboard *Einstein* inspired the calculation of X-ray line parameters (see, e.g., Kallman & McCray 1982) that later provided a rich and essential resource when *Chandra* and *XMM-Newton* launched with the first dispersive X-ray spectrometers. Current accretion disk simulations - and those conducted in the next 10 years - will provide a similar framework for understanding IXO observations of disk accretion onto compact objects.

**Starving Black Holes**

Active accretion phases are short-lived (e.g. Martini 2003, Hopkins et al. 2005; see Figure 2). A full understanding of the co-evolution of black holes and galaxies, then, requires that we understand how black holes interact with their local environments when mass only accretes at low fractions of the Eddington limit ($10^{-4}$, and below). Advanced theoretical studies of galaxy evolution suggest that *even within such "quiet" phases, feedback from black holes is required to match the observed colors of galaxies* (e.g. Croton et al. 2006). Robust physical arguments and simple observations show that a standard thin accretion disk cannot be the driving force when rates of mass transfer are very low. However, owing to the low luminosity of inefficient accretion and to the limited mirror area of current missions, astrophysicists are left to ask:

- **What is the fate of gas falling onto a black hole at a low rate?**

*In the simplest terms, we do not know if gas remains bound to a black hole at low accretion rates.* The paradigm for accretion onto black holes at low rates of mass transfer is that radiation is trapped within the flow, and *advected* across the event horizon with the gas (see, e.g., Narayan & Yi 1994). However, an independent treatment of the problem gives a very different answer: most accreting gas *must* be driven away in a wind due to viscous heating (Blandford & Begelman 1999). Winds are detected through line spectroscopy, and the temperatures and sizes of advective flows are revealed through emission lines. Current missions lack the sensitivity needed to detect lines. The high collecting area of IXO, its energy broad energy range, and superior instrumental capacities mean that astrophysicists will finally be able to make paradigm-testing observations of the accretion modes that dominate the lifetimes of black holes and their host galaxies.

The specific example of a modest survey can help to illustrate the major progress that improved spectroscopy will make. Stellar dynamics and gas dynamics have been used to constrain the mass of black holes in approximately 40 nearby galaxies. These black holes are the basis of the well-known M-sigma relationship between black hole mass and velocity dispersion (see, e.g., Tremaine et al. 2002). These galaxies generally lie within 30 Mpc, and most of the black holes have X-ray luminosities of $10^{40}$ erg/s or less. Investigating black holes with known masses would be particularly expedient, as accretion flow properties can be studied as a function of Eddington luminosity.

In exposures of 100 ksec or less, IXO can detect line spectra from each of these black holes, even if the plasma is extremely hot (up to $10^9$ Kelvin) and line-poor (see Figure 3). A wealth of physical information will be available for the first time: line ratios can be used to measure plasma temperatures, line strengths can be used to estimate the size of the accretion flow (Narayan & Raymond 1999), and line shifts will indicate whether gas is bound to the black hole. Within the context of this example survey, the nature of the flows can be determined as a function of black hole mass, mass accretion rate, and a host of feedback parameters such as the star formation rate.

Galactic compact objects will provide an important point of comparison. Transient Galactic sources evolve through orders of magnitude in mass accretion rate and luminosity. If, for instance, gas is bound to black holes at low accretion rates but mostly lost in a wind at *very* low accretion rates, this evolution can be revealed through IXO observations of transient Galactic sources. In a given year, several stellar-mass black hole and one neutron star transients can be expected to go into outburst and then decay. By also observing neutron star transients, the role of a hard stellar surface (versus an event horizon) in shaping accretion and outflows can also be addressed. For both black holes and neutron stars at typical distances, a 100 ksec IXO observation can detect line spectra at source luminosities of $10^{-6}$ Eddington, and below.

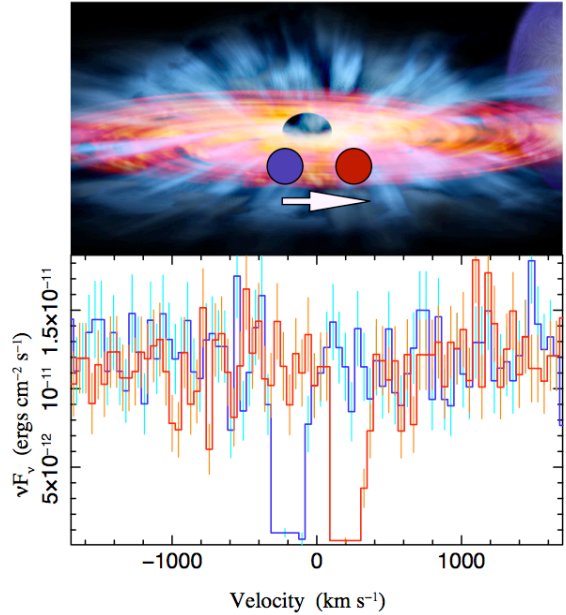

**Figure 1**: Strong tests of accretion physics become possible when orbital motion can be detected. IXO has the sensitivity to detect Doppler shifts in orbiting winds that can carry away angular momentum and permit disk accretion in Seyfert AGN. The figure above shows O VIII absorption lines with expected Doppler shifts from simulated 5 ksec observations of a transit with IXO.

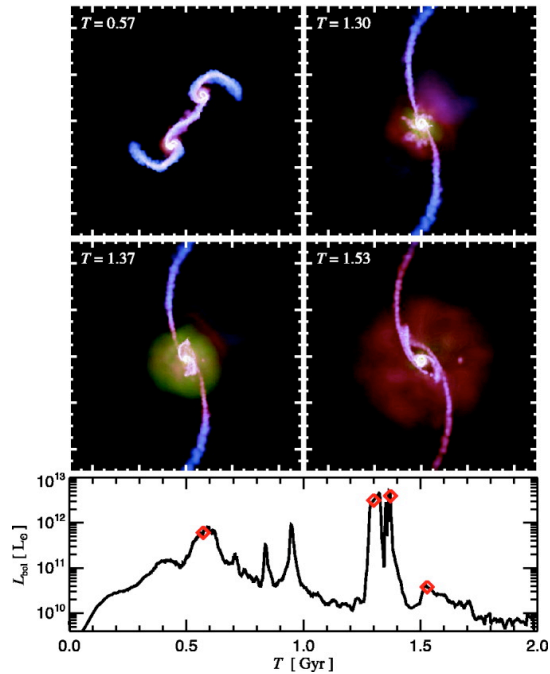

**Figure 2**: New simulations of galaxy mergers show that merger activity drives episodes of active accretion onto supermassive black holes. As shown above, active episodes are brief (Hopkins et al. 2005). To match the galaxy colors, however, the black hole must still exert influence on its host galaxy during inactive periods. The superior spectral resolution and collecting area of IXO will make it possible to reveal the physics of accretion onto supermassive black holes at low rates of mass transfer.

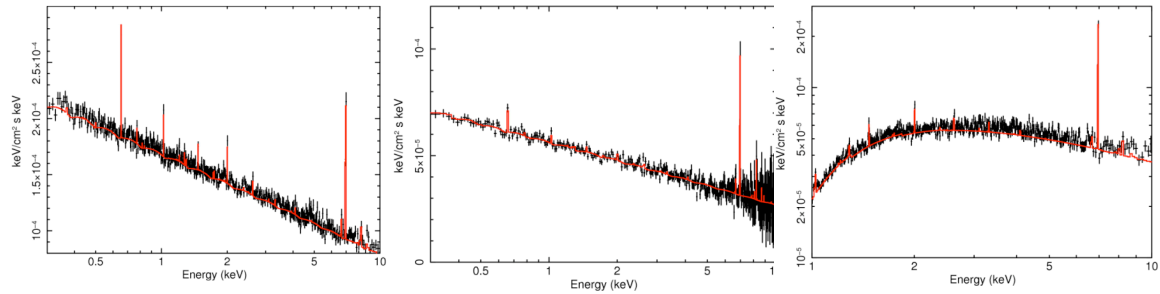

**Figure 3**: IXO will be able to detect lines from hot (50-100 keV) plasmas in a broad range of black holes, including those in nearby galaxies such as M87 (left), NGC 4143 (center), and stellar-mass black holes in the Milky Way such as V404 Cyg (right). With line spectra, paradigms for accretion onto black holes at low mass transfer rates can finally be tested.